# Marangoni Convection-Driven Laser Fountains and Waves on Free Surfaces of Liquids

Feng Lin[1,2], Aamir Nasir Quraishy[3,4], Tian Tong[2], Runjia Li[5], Guang Yang[6], Mohammadjavad Mohebinia[6], Yi Qiu[7,2], Talari Vishal[5], Junyi Zhao[8,2], Wei Zhang[9], Hong Zhong[1], Hang Zhang[1], Chaofu Zhou[1], Xin Tong[1], Peng Yu[1], Jonathan Hu[9], Suchuan Dong[10], Dong Liu[5], Zhiming Wang[1,*], John R. Schaibley[3,*], Jiming Bao[2,6,11,*]

1. Institute of Fundamental and Frontier Sciences, University of Electronic Science and Technology of China, Chengdu, Sichuan 610054, China
2. Department of Electrical and Computer Engineering, University of Houston, Houston, Texas 77204, USA
3. Department of Physics, University of Arizona, Tucson, Arizona 85721, USA
4. James C. Wyant College of Optical Sciences, University of Arizona, Tucson, Arizona 85721, USA
5. Department of Mechanical Engineering, University of Houston, Houston, Texas 77204, USA
6. Materials Science and Engineering Program, University of Houston, Houston, Texas 77204, USA
7. School of Science, Southwest Petroleum University, Chengdu, Sichuan 610500, China
8. Department of Electrical and Systems Engineering, Washington University in St. Louis, St. Louis, Missouri 63130, USA
9. Department of Electrical & Computer Engineering, Baylor University, Waco, Texas 76798, USA
10. Department of Mathematics, Purdue University, West Lafayette, Indiana 47907, USA
11. Department of Physics and Texas Center for Superconductivity, University of Houston, Houston, Texas 77204, USA

*To whom correspondence should be addressed. Emails: johnschaibley@email.arizona.edu, zhmwang@uestc.edu.cn, jbao@uh.edu.

# Abstract

It is well accepted that an outward Marangoni convection from a low surface tension region will make the surface depressed. Here, we report that this established perception is only valid for thin liquid films. Using surface laser heating, we show that in deep liquids a laser beam actually pulls up the fluid above the free surface generating fountains with different shapes. Whereas with decreasing liquid depth a transition from fountain to indentation with fountain-in-indentation is observed. Further, high-speed imaging reveals a transient surface process before steady elevation is formed, and this dynamic deformation is subsequently utilized to resonantly excite giant surface waves by a modulated laser beam. Computational fluid dynamics models reveal the underlying flow patterns and quantify the depth-dependent and time-resolved surface deformations. Our discoveries and techniques have upended the century–old perception and opened up a new regime of interdisciplinary research and applications of Marangoni-induced interface phenomena and optocapillary fluidic surfaces – the control of fluids with light.

Liquid surfaces are ubiquitous in our daily life. The Marangoni effect is the convection of fluid driven by a surface tension gradient. Marangoni-induced flow patterns, surface deformation and surface waves have been attracting enormous attention due to their fascinating complexity and enormous applications to dynamic fluid control(*1-4*). Since Marangoni convection pulls liquid away from the low tension region, it has been observed and widely accepted that the surface with a locally low tension will become depressed(*1-6*). Tears of wine and Bénard cells, driven by composition and temperature gradient, respectively(*5*), are good examples of Marangoni convection-driven surface depression although it took 50 years to identify the role of surface tension in Bénard cells because of potential contribution from natural convection(*7, 8*). The first quantitative description of surface depression was provided by Landau by considering the lateral flow only for a given surface tension gradient in a thin liquid film(*6*). More rigorous theoretical treatments(*9-16*) and controlled experiments(*10, 17-21*) were performed later with non-uniform surface tension created by either substrate heating or surface radiative heating. Besides these fundamental investigations, more cases have been investigated in enormous applications involving liquids or soft matters such as lithography and 3D printing(*22-24*), heat transfer and mass transport(*19*), crystal growth and alloy welding(*25, 26*), dynamic grating and spatial light modulator(*15, 27*), microfluidics and adaptive optics(*19, 28-30*). These studies further expand Landau's theory and strengthen previous qualitative understanding by providing a detailed description of surface depression and flow patterns.

Surface waves, *i.e.*, oscillatory deformations of the surfaces, can be regarded as excited states of surface, thus, they are intrinsic character of surface and can offer additional physics and potential applications compared to stationary surface deformation(*1-4, 6*). The discovery of the Marangoni effect provides us an alternative means to generate surface waves besides conventional mechanical methods(*1-3, 20, 21*). This new technique of surface wave generation has been demonstrated, however, only very small wave amplitude was achieved, e.g., on the order of micrometers with composition gradient(*2, 3, 31*) and substrate-induced temperature gradient(*2, 3, 31-33*), and nanometers with laser-induced temperature gradient(*20, 21*). Such small surface oscillations also require special techniques to measure, such as optical interference or reflection of a laser beam from the surface gradient(*2, 3, 20, 21, 31-33*), and therefore has limited the fundamental research and many potential applications of surface waves.

In this work we use a relatively low-power (<1.5 W) continuous-wave (CW) laser beam to create non-uniform surface temperature for the study of Marangoni effect. Compared with contact heating from liquid substrates or walls, surface laser heating has the advantage of being non-contact precise control of temperature in space and time without inducing natural convection, which is especially useful in a deep liquid. Figure 1a shows a schematic of our experimental setup. Ferrofluid is held in a petri dish and is illuminated from above with a 532-nm laser beam. The resulting distortions of the fluid surface are imaged with digital cameras. Commercial ferrofluid (educational ferrofluid EFH1 from Ferrotec Corporation) is chosen because of its high light-absorbing property for efficient surface heating. A cylindrical lens and axicon lens are used to generate a line beam and a ring beam, respectively, to generate different temperature patterns. We use a mechanical shutter and a mechanical chopper to control the illumination of laser.

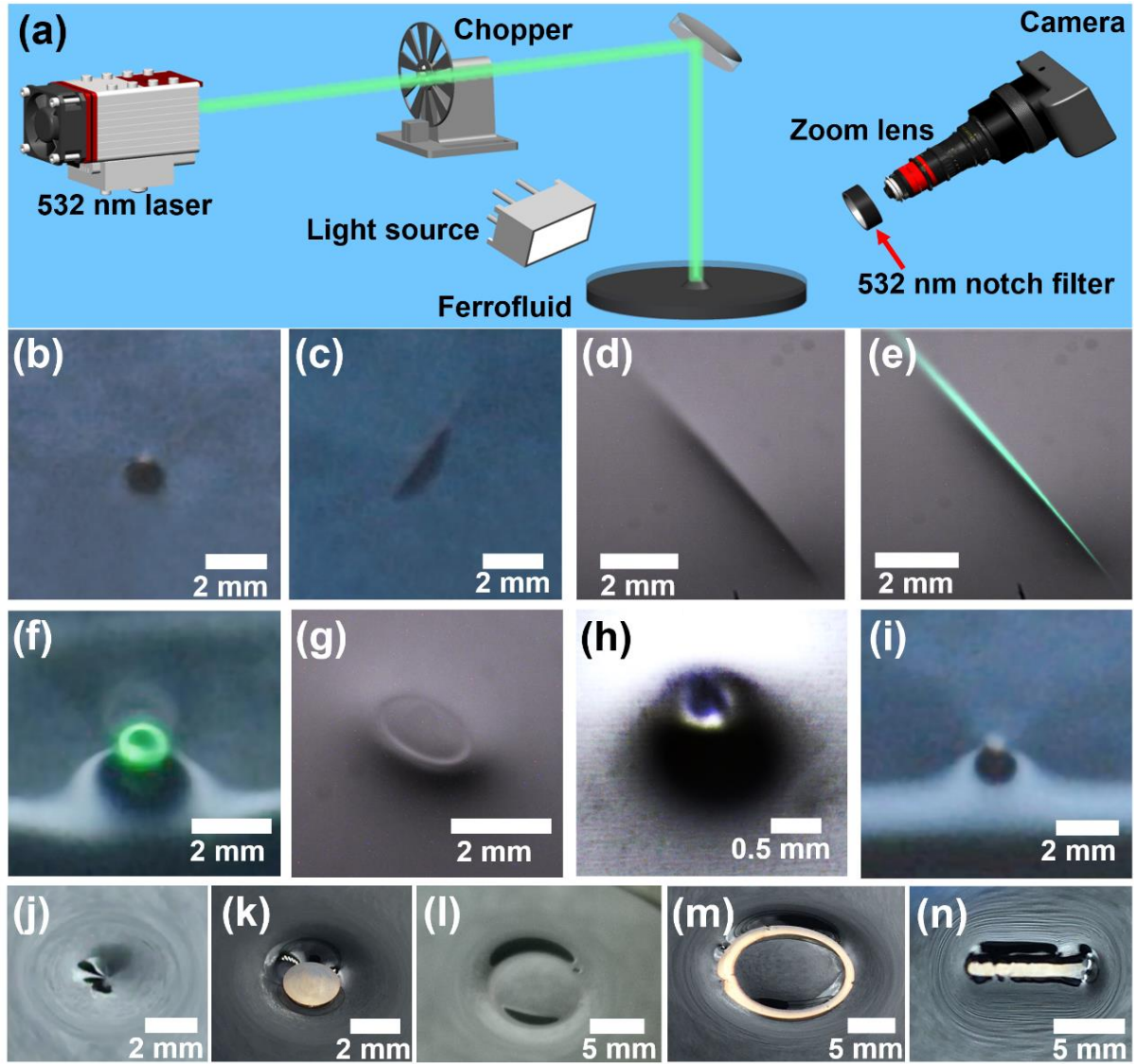

Figure 1. Experimental setup and images of stationary surface deformations in a ferrofluid generated by 532-nm continuous-wave laser beams. (a) Schematic of experimental setup. (b-h) Fountains with the shape of (b) a pillar, (c-e) a ridge and (f-h) a hat generated by a Gaussian beam, a line-focused beam and a ring beam, respectively. The pillar in (i) is generated by a highly focused ring beam. The depth of ferrofluid is 8 mm. (j-n) Surface indentations induced by the same shapes of laser beams in a shallow ferrofluid (0.5 mm). The laser is blocked in all except (e) and (f) by a notch filter. The laser power is 1 W or less than 1 W for (b) and (j-n), 2 W for (c-i).

Ferrofluid is a so-called "magic" liquid and is best known for its astonishing surface spikes generated by a magnet(*34*). Surprisingly, similar spikes can also be created by a laser. Figure 1b shows a spike or pillar on the flat surface of ferrofluid under a 0.6-mm diameter laser beam without any focusing. Like a magnetic field, the height of pillar increases with increasing laser power, and the pillar is stable as long as the laser is on. However, unlike a magnet, a laser beam can generate other surface protrusions by simply varying its shape. Figures 1c-h show examples of ridge and hat created by a line beam and a ring beam. A higher pillar can be created by a tightly focused ring beam (Figure 1i). Because their shapes can be arbitrarily controlled by the shape of laser beams, we call these surface protrusions "laser fountains." Like a mechanically pumped fountain, once the laser as a pump is turn off, the fountain will shut off in less than 0.1 second. It should be noticed that when the liquid layer is shallow, the same laser beams actually create different indentations (Figures 1j-n) even for the same ferrofluid. A higher laser power or a shallower layer will make depression so strong that the entire liquid layer become ruptured (Figs. 1k, 1m and 1n)(*2, 3, 10, 14*). We note that the magnitude of this effect is qualitatively different than previously reported surface depressions, which were too weak to be directly observed by the naked eye or ordinary photography(*10, 17-21*).

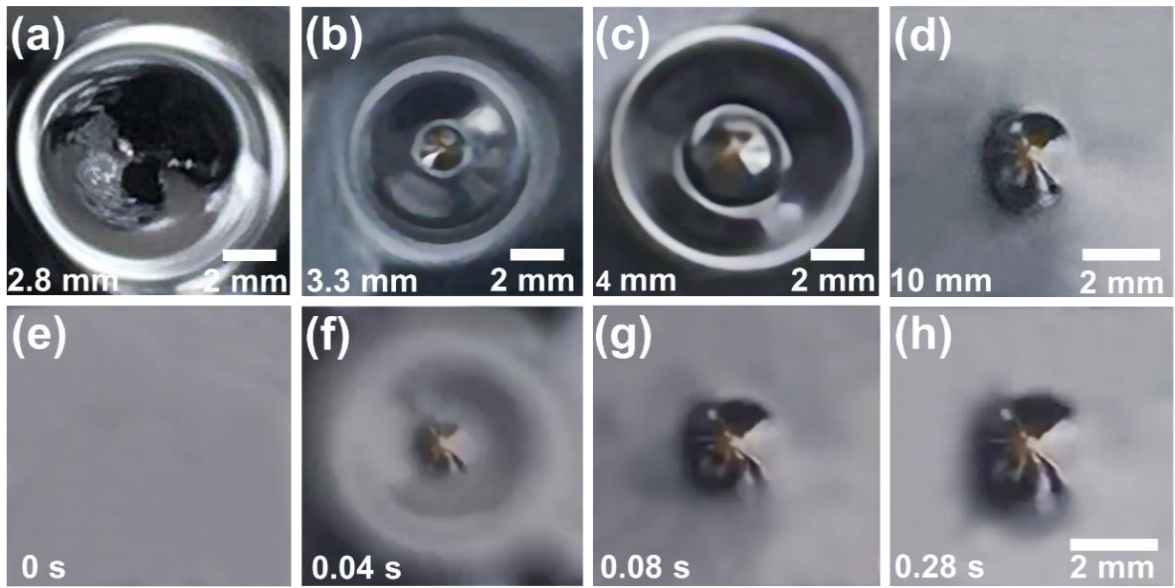

Figure 2. Images of depth-dependent and time-resolved surface deformations. (a-d) Emergence of a pillar fountain from the center of a larger indentation with increasing depth of ferrofluid from 2.8 mm in (a) to 10 mm in (d). (e-h) Time-resolved surface deformation in a deep ferrofluid (8 mm) after the turn-on of the laser showing initial formation of indentation followed by the growth and dominance of a pillar. The power of laser is 1 W.

To understand these distinct deformations between deep and shallow liquids, we varied the liquid layer thickness while keeping the laser beam the same. Figs. 2a-d show that an indentation is observed when the thickness of ferrofluid is 2.8 mm, in agreement with the observation made for a shallower ferrofluid layer in Figs. 1j-n as well as with the findings in past studies; however, in a 3.2-mm thick liquid, a spike begins to emerge from the center of the indentation. With increasing thickness, the spike continues to grow and the indentation becomes less and less visible. When the liquid thickness reaches 8 mm or above, the spike dominates the surface and the indentation disappears completely, *i.e.*, a laser fountain is formed. The laser fountains and the depth-dependent transition from surface indentation to laser fountain pattern have never been reported in the literature even in a deep fluid, most probably because they have not been predicted or investigated by any theory.

Despite the drastic difference in surface deformation for a shallow liquid and a deep liquid, the laser induced Marangoni convection, particularly, its divergent nature of outward convection from the laser spot is expected to be the same since laser heating will cause the local decrease in surface tension. To confirm this, we added fluorescent tracer microspheres to probe the Marangoni convection(*35*), and we found the same outward flow from the laser point in all the cases (with a representative one in the Supplementary Video). To understand why similar divergent Marangoni flow creates totally different surface deformation, we used a high-sped camera to probe the dynamics of surface profile right after the laser illuminates a deep ferrofluid. Time resolved images Figs. 2e-h clearly capture the draining of fluid and the formation of a depression due to the Marangoni convection before the appearance and slow growth of a spike. Thus, there is no simple relationship between the outward Marangoni convection and surface depression as previously understood.

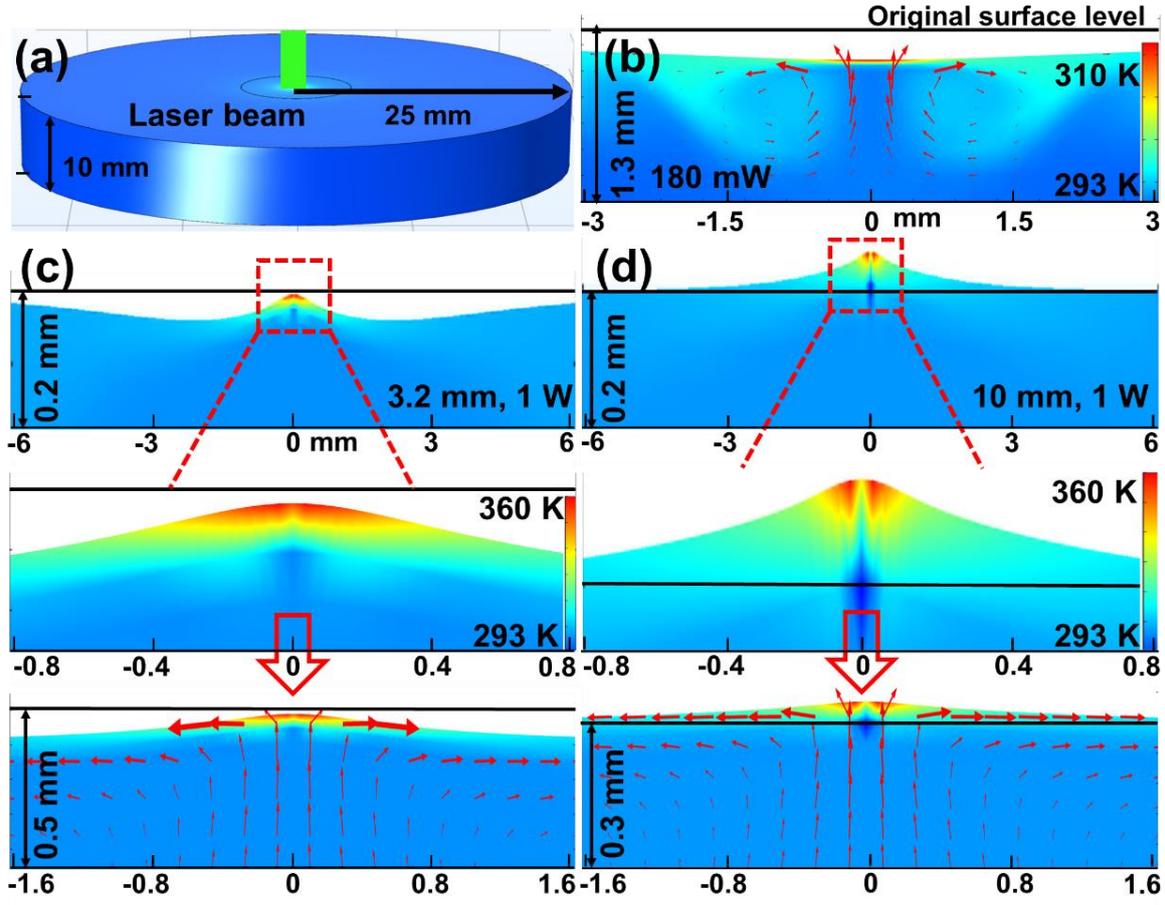

Figure 3. Stationary surface deformations, temperature distributions and flow patterns from COMSOL simulations. (a) Simulation configuration with a 10-mm thick liquid. The diameter of liquid is fixed at 5 cm. (b-d) Simulated surface deformation, flow and temperature fields in ferrofluid with depth of (b) 1.3 mm, (c) 3.2 mm and (d) 10 mm. (c) and (d) show selected areas for better viewing.

We emphasize that there have been numerous attempts to understand the Marangoni flow-driven surface deformation, but no existing theory can predict the deformation patterns of a liquid with an arbitrary depth in a straightforward manner. Approximations have to be often made even in analyzing surface deformation of a shallow liquid, due to complexities in solving the highly nonlinear Navier-Stokes equations, especially when the flow is driven by surface tension gradient and coupled with heat transfer and optical absorption, and the free surface – the boundary of the fluid, is not fixed. To avoid the limitations of oversimplified analytical models, we turned to computational fluid dynamics (CFD), which allows us to solve the complete optical-thermal-flow problem with changing boundary. Note that although ferrofluid is a composite nanofluid that is made of superparamagnetic nanoparticles suspended in hydrocarbon oil, it does not exhibit any magnetism in the absence of an external magnetic field and can still be well modeled as a homogeneous single-phase liquid(*34*). Figure 3a shows the COMSOL model of ferrofluid, which assumes a cylindrical symmetry under a vertically incident Gaussian laser beam. CFD simulation results in Figure 3b shows the expected Marangoni convection and surface indentation in a representative 1.3-mm-thick liquid, the simulation also confirms that the depression increases with decreasing thickness of the liquid layer until the rupture of the liquid – a manifestation of Marangoni instability. The failure of Landau's lateral flow approximation can also be seen from the vertical upward flow below the laser heated region.

Inspired by the successful simulation of surface depression in a shallow liquid, we increase the depth of ferrofluid in the simulation. Surprisingly, as Figures 3c and 3d show that the effect of liquid depth on the surface deformation can also be well predicted. A close comparison of flow patterns in the shallow and deep liquids reveals that the surface deformation is not induced by the outward Marangoni convection alone, it is actually determined by the competition between the Marangoni convection on the surface and the upward recirculation flow that comes from the subsurface colder region. In a shallower liquid as in Fig. 3b, the recirculation flow is hindered by viscous shear force from the substrate, so a depression is formed. As the liquid depth increases, there will be less and less flow resistance from the substrate, thus leading to a weaker depression. For a sufficiently deep liquid, the upward recirculation flow may become strong enough to outweigh the outward Marangoni flow, causing a fountain to form. It can also be seen that due to this strong upward cold flow from the bottom, the center spot on the free surface does not have the highest temperature.

Understanding the distinct surface deformation in liquids with different depths helps us to unravel the dynamics of the surface deformation process: the observed transient deformation must originate from a dynamic instead of a stationary competition between the Marangoni convection and the recirculation flow. This understanding is confirmed by the CFD simulations. Figures 4a-b show that, initially, a hot spot exists near the liquid surface (owing to slow heat diffusion) that result in the outward Marangoni convection from the laser beam center. Since the capillary flow is highly localized, there is only a very weak recirculation flow and the surface becomes depressed as observed in a shallow liquid. However, as shown in Figures 4c-d, the recirculation flow develops very quickly and the resulting convection heat transfer causes the surface temperature gradient to decrease. Eventually, the liquid mass carried by recirculation exceeds that by the outward Marangoni flow, leading to the formation of a spike in the center of indentation and ultimate stationary pillars above the flat surface.

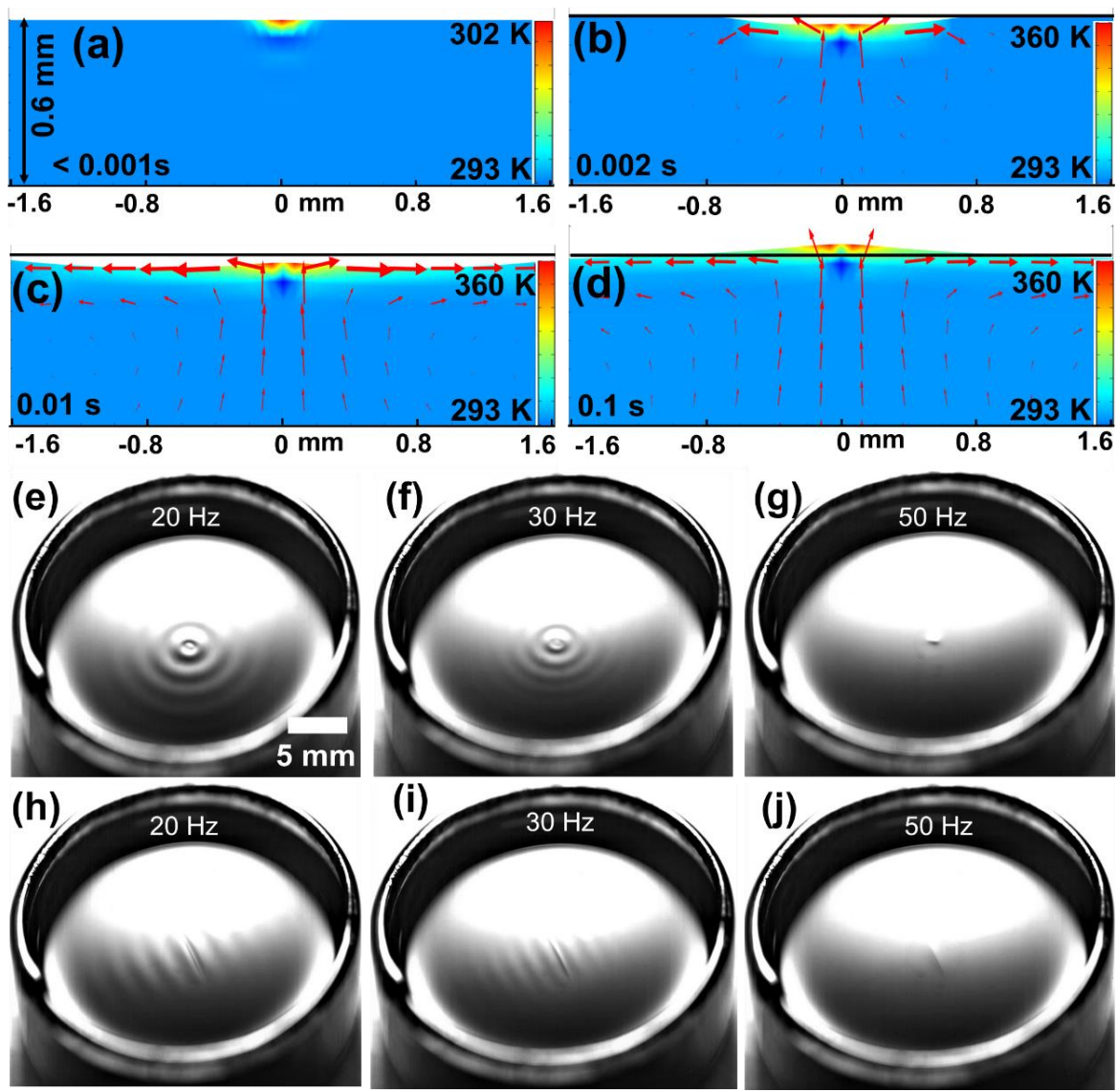

Figure 4. Simulated transient surface deformation and excitation of surface waves by modulated laser beams. (a-d) Temporally resolved simulated surface profile, temperature distribution and flow velocity in a deep ferrofluid. (e-j) Dependence of surface waves on laser modulation frequency for (e-g) circular waves and (h-j) near-plane waves in a deep ferrofluid (8 mm).

Our observation of oscillatory surface deformation, *i.e.*, a transient surface depression followed by elevation, motivates a new technique to generate surface waves. If we modulate the laser beam at a frequency that matches the dynamic surface depression-elevation cycle, the resulting surface wave amplitude can be greatly enhanced. To demonstrate this effect, we used a mechanical chopper to modulate the intensity of the laser and used a high-speed camera to image the surface wave. A giant surface wave can be seen for the first time in figure 4e at a modulation frequency of 20 Hz. The resonant excitation can be seen in Figs. 4f-g: the wave gets weaker at 30 Hz and eventually becomes invisible at 50 Hz. A near-plane wave can also be generated by a line-focused laser beam with a similar resonance, as shown in Figs. 4h-j.

In conclusion, we have generated laser fountains with various shapes through laser-induced thermocapillary force, we have observed and explained the depth-dependent and time-resolved surface deformations with both experiments and CDF simulations. We have also demonstrated a new way to excite strong surface wave. Since Marangoni effect happens in any liquid with an arbitrary depth, our discoveries, experimental and simulation techniques have unraveled previously limited understanding of Marangoni convection-induced interface phenomena and enabled us to explore new regimes from thin liquid films to a liquid with an arbitrary depth. Ferrofluid is not a unique liquid for Marangoni convection, a liquid's strong optical absorption can be achieved by varying the laser wavelength, or adding light absorbing elements such as colorants or plasmonic nanoparticles(*18, 35*). Besides fundamental research, Marangoni effect has already found enormous applications in nearly all branches of engineering due to the ubiquitous involvement of liquids(*19, 28-30*), the optical manipulation of fluids and fluid surface through Marangoni effect is expected to open new opportunities for novel applications.